%% file: BAD-1789.tex
\documentclass[twocolumn,showpacs,aps,prl,superscriptaddress]{revtex4}

\usepackage{graphicx}
\usepackage{dcolumn}
\usepackage{epsfig}
\usepackage{psfrag}
\usepackage{amsmath}

\newcommand{\BaBarYear}    {07}
\newcommand{\BaBarNumber}  {037}

\newcommand{\SLACPubNumber} {12700}

 \newcommand{\BaBarType}      {PUB}  

\input pubboard/babarsym   

\input Q2BDefn/Definitions.tex

\newcommand{\dw}{\ensuremath{\Delta w}}
\newcommand{\dC}{\ensuremath{\Delta C}}

\newcommand{\Rbzpip}{\ensuremath{(\rbzpip)\times 10^{-6}}}
\newcommand{\rbzpip}{\ensuremath{6.7\pm 1.7\pm 1.0}}
\newcommand{\sbzpip}{\ensuremath{4.0}}
\newcommand{\abzpip}{\ensuremath{0.05\pm 0.16\pm 0.02}}

\newcommand{\RbzKp}{\ensuremath{(\rbzKp)\times 10^{-6}}}
\newcommand{\rbzKp}{\ensuremath{9.1\pm 1.7\pm 1.0}}
\newcommand{\sbzKp}{\ensuremath{5.3}}
\newcommand{\abzKp}{\ensuremath{-0.46\pm 0.20\pm 0.02}}

\newcommand{\Rbmppipm}{\ensuremath{(\rbmppipm)\times 10^{-6}}}
\newcommand{\rbmppipm}{\ensuremath{10.9\pm 1.2\pm 0.9}}
\newcommand{\sbmppipm}{\ensuremath{8.9}}
\newcommand{\abmppipm}{\ensuremath{-0.05\pm 0.10\pm 0.02}}
\newcommand{\Cbmppipm}{\ensuremath{-0.22\pm 0.23\pm 0.05}}
\newcommand{\dCbmppipm}{\ensuremath{-1.04\pm 0.23\pm 0.08}}

\newcommand{\RbmKp}{\ensuremath{(\rbmKp)\times 10^{-6}}}
\newcommand{\rbmKp}{\ensuremath{7.4\pm 1.0\pm 1.0}}
\newcommand{\sbmKp}{\ensuremath{6.1}}
\newcommand{\abmKp}{\ensuremath{-0.07\pm 0.12\pm 0.02}}

\newcommand{\theTitle}{{\boldmath Observation of \B-meson decays to
$b_1\pi$ and $b_1 K$}} 

\begin{document}


\begin{flushleft}
\babar-\BaBarType-\BaBarYear/\BaBarNumber \\
SLAC-PUB-\SLACPubNumber \\
\end{flushleft}


\title{\theTitle}

\input pubboard/authors_jun2007.tex

\date{\today}

\begin{abstract}
We present the results of searches for decays of \B\ mesons to final
states with a \bone\ meson and a charged pion or kaon.  The data,
collected with the \babar\ detector at the Stanford Linear Accelerator
Center, represent 382 million \BB\ pairs produced in \epem\
annihilation.  The results for the branching fractions
are, in units of $10^{-6}$,
\Bbzpip = \rbzpip\ ($\sbzpip\sigma$),
\BbzKp = \rbzKp\ ($\sbzKp\sigma$),
\Bbmppipm  =  \rbmppipm\ ($\sbmppipm\sigma$), and
\BbmKp  =  \rbmKp\ ($\sbmKp\sigma$),
with the assumption that ${\cal B}(\bone\ra\omega\pi)=1$.
We also measure charge and flavor asymmetries
\acp(\bzpip) = \abzpip,
\acp(\bzKp) = \abzKp,
\acp(\bmppipm) = \abmppipm,
$C(\bmppipm) = \Cbmppipm$,
\dC(\bmppipm) = \dCbmppipm, and
\acp(\bmKp) = \abmKp.
The first error quoted is statistical, the second systematic, and for
the branching fractions, the significance is given in parentheses. 
\end{abstract}


\maketitle


Recent searches for decays of \B\ mesons to final states with an
axial-vector meson and a pion have revealed modes with rather large
branching fractions, e.g., ${\cal B}(\Bz\ra
a_1^\mp\pi^\pm)=(33.2\pm3.8\pm3.0)\times10^{-6}$ \cite{BaBar_a1pi}.
Here we search for related modes with a $b_{1}^{0}$ or a \bonem\ meson
plus a \pip\ or \Kp\ \cite{conjugate}, in a sample of
$(381.8\pm4.2)\times 10^6$ \BB\ pairs produced by \epem\ annihilation at
the $\Upsilon (4S)$ resonance (center-of-mass energy $\sqrt{s}=10.58\
\gev$).  The integrated luminosity is 346 \invfb.

The mass and width of the \bone\ are $1229.5\pm 3.2$ MeV and $142\pm 9$ MeV,
respectively, and the dominant decay is to $\omega\pi$ \cite{PDG2006}.
In the quark model the $b_1$ is the $I^G=1^+$ member of the
$J^{PC}=1^{+-},\ \relax^1\!P_1$ nonet, whereas the $a_1$ is the
$I^G=1^-$ state in the $J^{PC}=1^{++},\ \relax^3\!P_1$ nonet.
The available theoretical estimates of the branching fractions of \B
mesons to $\bone\pi$ and $\bone K$ come from calculations based on naive
factorization \cite{laporta,calderon}, and on QCD factorization
\cite{ChengYang}.  The latter incorporate light-cone distribution
amplitudes evaluated from QCD sum rules.  Expected branching fractions
lie in the range 5--10$\times10^{-6}$ \cite{ChengYang}; estimates
as large as 26$\times10^{-6}$ are found in the calculations of
\cite{laporta}, and 40$\times10^{-6}$ in those of \cite{calderon}.

The four modes \bzpip, \bzKp, \bmpip, and \bmKp\ can be mediated by
external tree amplitudes in which the weak current produces the pion
(kaon) with a Cabibbo-favored (suppressed) coupling.  Alternatively, a
``penguin'' loop amplitude is favored for the kaon modes, and
suppressed for the pion modes.  The fifth mode, \bppim, requires a coupling of
the current to the \bonep, which is forbidden for this $G=+1$ state
\cite{weinberg}, leading to the expectation $\Bbppim\ll\Bbmpip$.

Direct \CP\ violation would be indicated by a non-zero value of
the asymmetry $\acp \equiv (\Gamma^--\Gamma^+)/(\Gamma^-+\Gamma^+)$ in
the rates $\Gamma^\pm(B^\pm\ra F^\pm)$ for decay of a charged \B\ meson,
or  $\Gamma^+(\bmKp)$ and its charge conjugate. 
For the decays \bmppipm\ we define \acp\ and two additional
asymmetries $C$ and \dC\ through
\beq
\Gamma_{q,f} = \frac{1}{4}(\Gamma+{\overline\Gamma}\xspace)
 \left(1+q\acp\right)\left[1+f\left(C+q\dC\right)\right],
\eeq
where the signal $B$ meson flavor $f=+1$ for \Bz, $-1$ for \Bzb, and
$q$ is the sign of charge of the \bone.
To measure $C$ and \dC\ we use
the flavor $\eta\ (+1$ for \Bz\ and $-1$ for \Bzb) of the second meson
$B_{\rm tag}$ produced in \UfourS\ decay \cite{bbBBbarTag}.  The
yields are given by 
\begin{eqnarray}\label{eq:Ycharge}
Y_{q\eta}&=&\frac{1}{4}Y_S\left(1+q\acp\right)
\Big\{
1-\eta\dw+\eta\mu(1-2w)\Big.\\ 
&&\!\!\!\!\Big.-\eta(1-2\chi_d)\left[1-2w+\mu(\eta-\dw)\right](C+q\dC)\Big\}\,,\nonumber
\end{eqnarray}
where $Y_S$ is the total signal yield, $\chi_d=0.188\pm0.003$ the
time-integrated mixing probability \cite{PDG2006}, 
$w$ the mistag fraction, and \dw\ and $\mu$ the 
$B-\Bbar$ differences in the mistag rate and tagging efficiency,
respectively.

The data were collected with the \babar\
detector~\cite{BABARNIM} at the PEP-II asymmetric $e^+e^-$
collider located at the Stanford Linear Accelerator Center.
Charged particles from the \epem\ interactions are detected, and their
momenta measured, by a combination of five layers of double-sided
silicon microstrip detectors and a 40-layer drift chamber, both
operating in the 1.5~T magnetic field of a superconducting
solenoid. Photons and electrons are identified with a CsI(Tl)
electromagnetic calorimeter (EMC).  Further charged particle
identification (PID) is provided by the average energy loss ($dE/dx$) in
the tracking devices and by an internally reflecting ring imaging
Cherenkov detector (DIRC) covering the central region.  A detailed
Monte Carlo program (MC) is used to simulate the \B
production and decay sequences, and the detector response
\cite{geant}.

\begin{figure*}
\psfrag{FFF}{{$\cal F~~~~$}}
\includegraphics[width=1.\linewidth]{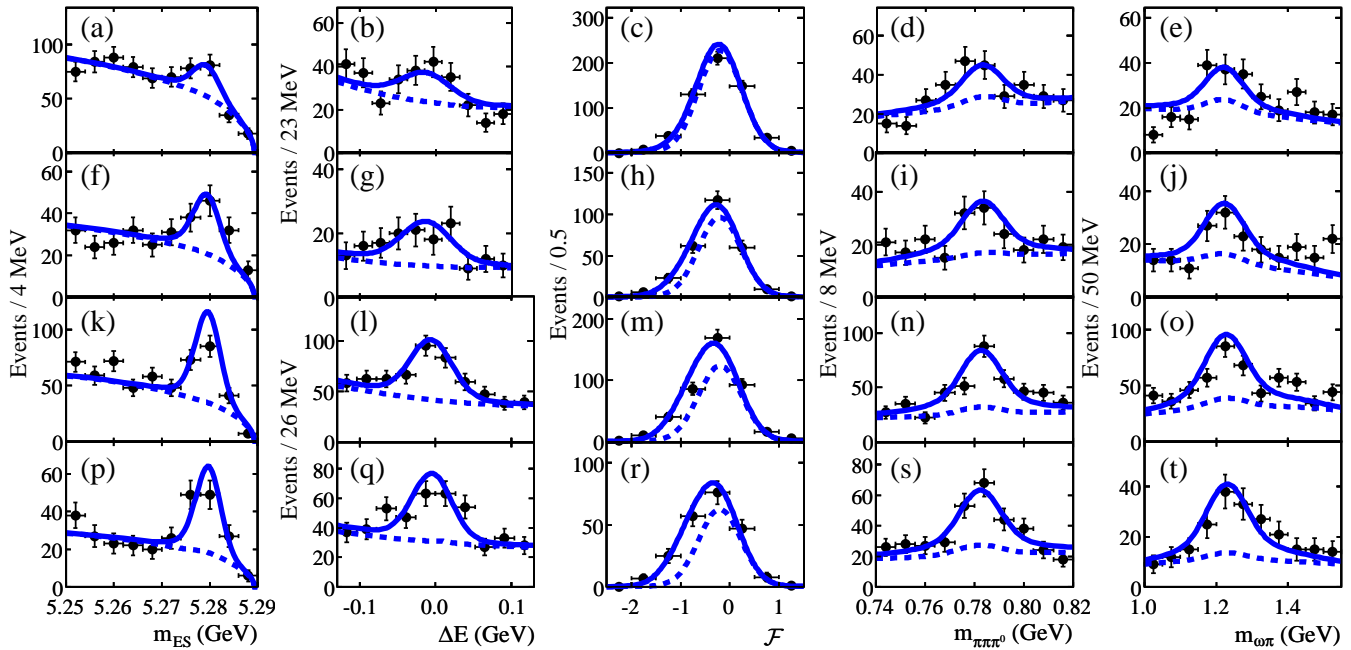}
\caption{\label{fig:proj_all}
Distributions for signal-enhanced subsets
of the data projected onto the fit observables for the decays:  (a-e)
\bzpip, (f-j) \bzKp, (k-o) \bmppipm, and (p-t) \bmKp.  The solid line
represents the result of the fit, and the dashed line the background
contribution.
}
\end{figure*}

The \bone\ candidates are reconstructed through the decay sequence
$b_1\ra\omega\pi$, $\omega\ra\pip\pim\piz$, and $\piz\ra\gaga$.  The
invariant mass of the photon pair is required to lie between 
120 and 150 MeV, i.e., within about two standard deviations of the nominal
mass \cite{PDG2006}.  For the \bone\ and $\omega$ whose masses are
observables in the maximum likelihood (ML) fit described below, we accept
a range that includes wider 
sidebands (see Fig.\ \ref{fig:proj_all}).  Secondary charged pions in
\bone\ and $\omega$ candidates are rejected if classified as protons,
kaons, or electrons by their DIRC, $dE/dx$, and EMC PID signatures.  For
the primary pion (kaon) from the \B-meson decay we define the PID
variable $S_{\pi}$ ($S_K$) as the number of standard deviations
between the measured DIRC Cherenkov angle and that expected for a pion
(kaon), requiring $-2<S_\pi<5$ ($-5<S_K<2$).

We reconstruct the \B-meson candidate by combining the 4-momenta of
a pair of daughter mesons, using a fit that constrains all particles to
a common vertex and the \piz\ mass to its nominal value.  From the
kinematics of \UfourS\ decay we determine the 
energy-substituted mass $\mes=\sqrt{\frac{1}{4}s-\pvec_B^2}$ and
energy difference $\DE = E_B-\half\sqrt{s}$, where $(E_B,\pvec_B)$ is
the \B-meson 4-momentum vector, and all values are expressed in the
\UfourS\ rest frame.  The resolution in \mes\ is $2.4-2.7\ \mev$ and in
\DE\ is 25--32 MeV, depending on the decay mode.  We require $5.25\
\gev<\mes<5.29\ \gev$ and $-0.13\ \gev<\DE<\DE_{\rm max}$, with
$\DE_{\rm max}=0.1\ (0.13)\ \gev$ for \bonez\ (\bonep), where the
tighter restriction serves to limit the number of combinatorial
candidates per event.  

We also impose restrictions on resonance decay angles to exclude the
most asymmetric decays where soft-particle backgrounds accumulate and
the acceptance changes rapidly.  We require 
$\cos\theta_{\bone} \le 1.1 - 0.5|\cos\theta_{\omega}|$, where 
$\theta_{\bone}$ is the angle
between the momenta of the pion from $\bone\ra\omega\pi$ and its
parent \B\ meson, measured in the $\bone$ rest frame, and
$\theta_{\omega}$ is the angle
between the normal to the $\omega\ra3\pi$ decay plane and the momentum
of its parent \bone, measured in the $\omega$ rest frame.  
Backgrounds arise primarily from random combinations of particles in
continuum $\epem\ra\qqbar$ events ($q=u,d,s,c$).  We reduce these with
a requirement on the angle \thetaT\ between the thrust axis of the \B
candidate in the \UfourS\ frame and that of the rest of the charged
tracks and neutral calorimeter clusters in the event.  The
distribution is sharply peaked near $|\costhr|=1$ for \qqbar\ jet
pairs, and nearly uniform for \B-meson decays.  The requirement,
which optimizes the expected signal yield relative to its
background-dominated statistical error, is $|\costhr|<0.7$.
The average number of candidates found per 
selected event is in the range 1.3 to 1.4 (1.4 to 1.6 in signal MC),
depending on the final 
state.  We choose the candidate with $\omega\pi$ invariant mass closest 
to the nominal value of the \bone\ mass \cite{PDG2006}. 
In the ML fit we discriminate further against \qqbar\ background with
a Fisher discriminant \xf\ that combines several variables which
characterize the energy flow in the event \cite{PRD04}.  It provides
about one standard deviation of separation between \B\ decay events
and \qqbar\ background.

We obtain yields for each channel from an extended ML
fit with the input observables \DE, \mes, \xf, and the resonance masses
$m_{\bone}$ and $m_\omega$.   
The selected data sample sizes are given in
Table~\ref{tab:results}.  Besides the signal events these samples
contain \qqbar\ 
(dominant) and \BB\ with $b\ra c$ combinatorial background, and a
fraction of cross feed from other charmless \BB\ modes, which we estimate
from the simulation to be (0.5--0.8)\%.  The last include non-resonant
$\omega\pi(\pi,K)$, and modes that have
final states different from the signal, but with similar
kinematics so that broad peaks near those of the signal appear in some
observables, requiring a separate component in the probability density
function (PDF).  The likelihood function is
\begin{eqnarray}
{\cal L} &=& \exp{\left(\smash[b]{-\sum_{j,q,\eta} Y_{j,q\eta}}\right)}
\prod_i^{N}\sum_{j,q,\eta} Y_{j,q\eta} \times \label{eq:likelihood}\\
&&{\cal P}_j (\mes^i) {\cal P}_j(\xf^i) {\cal P}_j (\DE^i) {\cal P}_j
(m_{\bone}^i) {\cal P}_j (m_{\omega}^i),
\nonumber  
\end{eqnarray}
where $N$ is the number of events in the sample, and for each
component $j$ (signal, combinatorial background, or charmless \BB\
cross feed), $Y_{j,q\eta}$ is the yield of events (Eq.\
\ref{eq:Ycharge}) and ${\cal P}_j(x^i)$ the PDF for observable $x$ in
event $i$.  The signal component is further  
separated into two components (with proportions fixed in the fit for each
mode) representing the 
correctly and incorrectly reconstructed candidates in events with true
signal, as determined with MC.  The factored form of the PDF
indicated in Eq.\ \ref{eq:likelihood}\ is a good approximation,
particularly for the combinatorial \qqbar\ component, since we find
correlations among observables in the data (which are mostly
\qqbar\ background) to be small.  The effects of this
approximation are determined in simulation and included 
in the bias corrections and systematic errors discussed below.

We determine the PDFs for the signal and \BB\ background components
from fits to MC samples.  We calibrate the resolutions in \DE\ and
\mes\ with large data control samples of \B\ decays to charmed final states
of similar topology (e.g.\ $B\ra D(K\pi\pi)\pi$).  We develop PDFs for
the combinatorial background with fits to the data from which the signal
region ($5.27\ \gev<\mes<5.29\ \gev$ and $|\DE|<0.1$ GeV) has been excluded.

The functions ${\cal P}_j$ are constructed as linear combinations of
Gaussian and 
polynomial functions, or in the case of \mes\ for \qqbar
background the threshold function
$x\sqrt{1-x^2}\exp{\left[-\xi(1-x^2)\right]}$, with argument
$x\equiv2\mes/\sqrt{s}$ and parameter $\xi$.  These functions are
discussed in more detail in \cite{PRD04}, and are illustrated in
Fig.~\ref{fig:proj_all}.

We allow the parameters most important for the determination of the
combinatorial background PDFs to vary in the fit, along with the
yields for all components, and the signal and
\qqbar\ background asymmetries.  Specifically, the free
background parameters are: $\xi$ for \mes, linear and quadratic
coefficients for \DE, and the mean, width, and width difference and
polynomial fraction parameters for \xf.

\input restab

We validate the fitting procedure by applying it to ensembles
of simulated experiments with the \qqbar\ component drawn from the PDF,
into which we have 
embedded the expected number of signal and \BB\ background events
randomly extracted from the fully simulated MC samples.  Biases obtained
by this procedure with inputs that reproduce the yields found in the
data are reported, along with the signal yields, in Table~\ref{tab:results}.

In Fig.\ \ref{fig:proj_all}\ we show the projections of the PDF and data
for each fit.  The data plotted are subsamples enriched in signal
with a threshold requirement on the ratio of signal to total likelihood
(computed without the plotted variable) that retains (29--53)\%
of the signal, depending on the mode.

We compute the branching fraction by subtracting the fit bias from the
measured yield, and dividing the result by the efficiency times
${\cal B}(\omega\ra\pip\pim\piz)=89.1\pm0.7\%$ \cite{PDG2006}, and by the
number of produced \BB\ pairs. 
We assume $\Gamma(\UfourS\ra\Bp\Bm)/\Gamma(\UfourS\ra\Bz\Bzb) = 1$,
consistent with measurements \cite{PDG2006}.
The results are given in Table~\ref{tab:results},  
along with the significance, computed as the square root of the difference
between the value of $-2\ln{\cal L}$ (with additive systematic
uncertainties included) for zero signal and the value at its minimum.

Systematic uncertainties on the branching fractions arise from the PDFs,
\BB\ backgrounds, fit bias, and efficiency.  PDF uncertainties not
already accounted for by free parameters in the fit are estimated from
the consistency of fits to MC and data in control modes.  Varying the
signal-PDF parameters within these errors, we estimate yield
uncertainties of (2.4--3.3)\%, depending on the mode.  The uncertainty
from fit bias (Table \ref{tab:results}) includes its statistical
uncertainty from the simulated experiments, and half of the correction
itself, added in quadrature.  For the \BB\ backgrounds we vary the
fixed fit component by 100\% and include in quadrature a term derived from 
MC studies of the inclusion of a $b\to c$ component with the dominant
\qqbar\ background.
Uncertainties in our knowledge of the efficiency
include $0.5\%\times N_t$ and $1.5\%\times N_\gamma$, where
$N_t$ and $N_\gamma$ are the numbers of tracks and photons, respectively,
in the \B\ candidate.  
The uncertainties in the efficiency from the event selection are
below 0.5\%.

We study asymmetries from the track reconstruction (found negligible),
and from imperfect modeling of the interactions with material in the
detector, by measuring the asymmetries in the \qqbar\ background in
the data and control samples mentioned previously, in comparison with
MC \cite{KpiDCP}.  We apply corrections, and assign systematic errors,
to \acp\ equal to $-0.010\pm0.005$ for modes with a primary kaon and
$0.000\pm0.005$ for those with a primary pion.  The leading systematic
errors on $C$ and \dC\ come from the fit bias.

With the assumption that ${\cal
B}(\bone\ra\omega\pi) = 1$, we obtain
for the branching fractions:
$$
\begin{array}{ccc}
\Bbzpip &=& \Rbzpip \\ 
\BbzKp &=& \RbzKp \\
\Bbmppipm  &=&  \Rbmppipm   \\
\BbmKp  &=&  \RbmKp.
\end{array}
$$
For the asymmetries we find
$$
\begin{array}{ccc}
\acp(\bzpip) &=& \abzpip \\
\acp(\bzKp) &=& \abzKp \\
\acp(\bmppipm) &=& \abmppipm  \\
C(\bmppipm) &=& \Cbmppipm  \\
\dC(\bmppipm) &=& \dCbmppipm  \\
\acp(\bmKp) &=& \abmKp. 
\end{array}
$$ 
The first error quoted is statistical and the second systematic.
The QCD factorization estimates \cite{ChengYang}\ for the branching
fractions and charge asymmetries agree with these measurements within
experimental and theoretical errors.  The authors of \cite{ChengYang}
note that the observation $\BbzKp/\BbmKp>0.5$, if confirmed with higher
precision, would indicate the presence of a weak annihilation
contribution to these modes.
The value of the
\CP-conserving \dC\ near $-1$ for \bmppipm\ agrees with the expected
suppression of \bppim; our results imply the ratio
$\Gamma(\bppim)/\Gamma(\bmppipm) = -0.01\pm0.12$.  We find no evidence
for direct \CP\ violation in these decays. 

\input pubboard/acknow_PRL

%

\renewcommand{\baselinestretch}{1}

\end{document}

%% file: Q2BDefn/Definitions.tex
\RequirePackage{xspace}

\newcommand{\pvec}{{\bf p}}

\newcommand{\acp}{\ensuremath{\calA_{ch}}}

\newcommand{\calB}{\ensuremath{{\cal B}}}

\newcommand{\DE}{\ensuremath{\Delta E}}

\newcommand{\xf}{\ensuremath{{\cal F}}}

\newcommand{\thetaT}{\ensuremath{\theta_{\rm T}}}
\newcommand{\costhr}{\ensuremath{\cos\thetaT}}

\newcommand\etal{{\it et al.}}
\newcommand{\half}{\ensuremath{\frac{1}{2}}}

\newcommand{\bfig}{\begin{figure}[htbpc!]}
\newcommand{\efig}{\end{figure}}
\newcommand\bef{\begin{figure}}
\newcommand\edf{\end{figure}}
\newcommand\dbline{\noalign{\vskip 0.10truecm\hrule}\noalign{\vskip 2pt}\noalign{\hrule\vskip 0.10truecm}}
\providecommand{\tbline}{\noalign{\vskip 0.05truecm\hrule\vskip0.05truecm}}

\newcommand\beq{\begin{equation}}
\newcommand\eeq{\end{equation}}
\newcommand\bear{\begin{array}}
\newcommand\enar{\end{array}}
\newcommand\beqa{\begin{eqnarray}}
\newcommand\eeqa{\end{eqnarray}}
\newcommand\ben{\begin{enumerate}}
\newcommand\een{\end{enumerate}}

\newcommand{\UfourS}{\ensuremath{\Upsilon(4S)}}

\newcommand{\bone}{\ensuremath{b_1}}
\newcommand{\bonep}{\ensuremath{b_1^+}}
\newcommand{\bonem}{\ensuremath{b_1^-}}
\newcommand{\bonemp}{\ensuremath{b_1^\mp}}

\newcommand{\bonez}{\ensuremath{b_1^0}}

\newcommand{\fbmpip}{\ensuremath{\bonem \pip}\xspace}
\newcommand{\bmpip}{\ensuremath{\Bz\ra\fbmpip}\xspace}
\newcommand{\Bbmpip}{\ensuremath{\calB(\bmpip)}\xspace}

\newcommand{\fbppim}{\ensuremath{\bonep \pim}\xspace}
\newcommand{\bppim}{\ensuremath{\Bz\ra\fbppim}\xspace}
\newcommand{\Bbppim}{\ensuremath{\calB(\bppim)}\xspace}

\newcommand{\fbmppipm}{\ensuremath{\bonemp \pipm}\xspace}
\newcommand{\bmppipm}{\ensuremath{\Bz\ra\fbmppipm}\xspace}
\newcommand{\Bbmppipm}{\ensuremath{\calB(\bmppipm)}\xspace}

\newcommand{\fbmKp}{\ensuremath{\bonem \Kp}\xspace}
\newcommand{\bmKp}{\ensuremath{\Bz\ra\fbmKp}\xspace}
\newcommand{\BbmKp}{\ensuremath{\calB(\bmKp)}\xspace}

\newcommand{\fbzpip}{\ensuremath{\bonez \pip}\xspace}
\newcommand{\bzpip}{\ensuremath{\Bp\ra\fbzpip}\xspace}
\newcommand{\Bbzpip}{\ensuremath{\calB(\bzpip)}\xspace}

\newcommand{\fbzKp}{\ensuremath{\bonez \Kp}\xspace}
\newcommand{\bzKp}{\ensuremath{\Bp\ra\fbzKp}\xspace}
\newcommand{\BbzKp}{\ensuremath{\calB(\bzKp)}\xspace}

%% file: pubboard/authors_jun2007.tex
%
\author{B.~Aubert}
\author{M.~Bona}
\author{D.~Boutigny}
\author{Y.~Karyotakis}
\author{J.~P.~Lees}
\author{V.~Poireau}
\author{X.~Prudent}
\author{V.~Tisserand}
\author{A.~Zghiche}
\affiliation{Laboratoire de Physique des Particules, IN2P3/CNRS et Universit\'e de Savoie, F-74941 Annecy-Le-Vieux, France }
\author{J.~Garra~Tico}
\author{E.~Grauges}
\affiliation{Universitat de Barcelona, Facultat de Fisica, Departament ECM, E-08028 Barcelona, Spain }
% ...
\collaboration{The \babar\ Collaboration}
\noaffiliation

%% file: restab.tex
\begin{table*}[btp]
\caption{
Number of events $N$ in the sample, fitted signal yield $Y_S$, and 
measured bias (to be subtracted from $Y_S$) in events (ev.), detection
efficiency $\epsilon$, 
significance~$\cal S$ (with systematic uncertainties included), and
branching fraction and charge asymmetry with
statistical and systematic error.
}
\label{tab:results}
\newcommand{\mn}{\ensuremath{\phantom{-}}}
\newcommand{\on}{\ensuremath{\phantom{1}}}
\newcommand{\eff}{$\epsilon$ (\%)}
\newcommand{\pbf}{$\prod\calB_i$ (\%)}
\newcommand{\signf}{$\cal S$ ($\sigma$)}
\begin{tabular}{lcr@{}lr@{}lccll}
\dbline
Mode	      	& $N$ (ev.)
			&\multicolumn{2}{c}{~$Y_S$ (ev.)~}
						&\multicolumn{2}{c}{~Bias (ev.)~}
									&\eff	
											&~~\signf
 													&\multicolumn{1}{c}{\calB\ $(10^{-6})$}	
															&\multicolumn{1}{c}{\acp}	\\
\tbline
\fbzpip		& 32176	& ~~~$178$&$^{+39}_{-37}$	& ~~~$26$&$\pm14$	& 6.78	& $\mn\sbzpip$	& $\on\rbzpip$	& \mn\abzpip	\\
\fbzKp		& 18036	&    $219$&$^{+38}_{-36}$	&    $24$&$\pm12$	& 6.73	& $\mn\sbzKp$	& $\on\rbzKp$	& \abzKp	\\
\fbmppipm	& 36901	&    $387$&$^{+41}_{-39}$	&    $34$&$\pm17$	& 9.54	& $\mn\sbmppipm$& $\rbmppipm$	& \abmppipm	\\
\fbmKp		& 17497	&    $267$&$^{+33}_{-32}$	&    $32$&$\pm16$	& 9.43	& $\mn\sbmKp$	& $\on\rbmKp$	& \abmKp	\\
\dbline
\end{tabular}
\end{table*}

%% file: pubboard/acknow_PRL.tex
We are grateful for the excellent luminosity and machine conditions
provided by our \pep2\ colleagues, 
and for the substantial dedicated effort from
the computing organizations that support \babar.
The collaborating institutions wish to thank 
SLAC for its support and kind hospitality. 
This work is supported by
DOE
and NSF (USA),
NSERC (Canada),
CEA and
CNRS-IN2P3
(France),
BMBF and DFG
(Germany),
INFN (Italy),
FOM (The Netherlands),
NFR (Norway),
MES (Russia),
MEC (Spain), and
STFC (United Kingdom). 
Individuals have received support from the
Marie Curie EIF (European Union) and
the A.~P.~Sloan Foundation.

%% file: BAD-1789.bbl
\begin{thebibliography}{99}
\bibitem{conjugate}
Charge-conjugate reactions are implied unless noted.
\end{thebibliography}
